\documentstyle[11pt,amstex]{article}
\begin{document}
\pagestyle{empty}
\begin{center} 
{\bf SOURCE DRIVEN SOLUTIONS}\\
{\bf OF QUANTUM FIELD THEORIES}\footnote{Preprint number: MIT-CTP-2577.  Supported in part by funds provided
by the U.S. Department of Energy (D.O.E.) under cooperative research agreement \#DF-FC02-94ER40818
and by Grant DE-FG09-91-ER-40588---Task D}\\
[10mm] 

G.S. GURALNIK\footnote{Permanent address: Department of Physics, Brown University, Providence, RI  02912, USA.  E-mail address: {\tt gerry@@het.brown.edu}}\\
[10mm]

{\em Center for Theoretical Physics\\
Laboratory for Nuclear Science\\
and Department of Physics\\
Massachusetts Institute of Technology\\
Cambridge, Massachusetts 02139}

\vspace{8mm}
\end{center}

\section{Introduction}

This talk describes a program to develop a new numerical approach to the solution of quantum field theories \cite{sg1} which we call the Source Galerkin method. 
The ideas involved have little directly in common with the Monte Carlo techniques that have demonstrated the power of numerical approaches to obtain results from quantum field theories. 
Source Galerkin has strengths which should allow its application to problems, such as scattering predictions, that require too much computing power to be examined by other methods at this time. 
Unlike Monte Carlo evaluations, Source Galerkin is equally suited for studying fermionic or bosonic problems. 
Further, at least in lower dimensions, a continuum formulation of Source Galerkin is straightforward. 
We have discovered that, while Source Galerkin implementations appear to be relatively economical in their use of computer cycles, they require complicated algebraic set up procedures for each problem examined. 
These take a good deal of care and large amounts of human time. 
While we have made considerable progress in understanding the issues in defining and applying these ideas, we have not yet been able to complete enough calculations to evaluate the chances that they might be able to supersede Monte Carlo techniques in any
 of their traditional applications.
We have, however, already obtained a striking result arising from our ideas. 
This result consists of a deep new insight into the solutions of QFT and phase structure \cite{zach1,zach2,gg1}.
As one aspect of this, we have shown that the conventional definition of the path integral does not include all solutions of canonical quantum field theory because a very restricted set of boundary conditions is implicitly assumed. 
The full solution set can be obtained by appropriate continuations of the integration regions into the complex plane.
This insight was a result of the very detailed and specific understanding of a theory that is necessary in order to numerically perform evaluations using Source Galerkin methods.

To apply the Source Galerkin ideas, we first write a field theory action in the presence of external sources for the fields. 
Then, by enlightened guessing, an approximate set of solutions to the linear source functional differential equations for the vacuum-vacuum amplitude ${\cal Z}$ of this action is constructed.
In principle, a terrible initial guess could be made. 
Through iteration this usually will eventually converge to a correct solution.
For a QFT it is possible and desirable to utilize symmetries expected of the exact answer in the construction of the approximate solutions.
This possibility is one of the great advantages of our method. 
In contrast it is difficult, or impossible, to insert knowledge obtained from analytic techniques into the construction of solutions using Monte Carlo techniques.

A straightforward way to make a Galerkin guess is to pick some linear combination of functions from a subset of a complete set of functions of the source variables for the system being studied. 
The use of a complete set of functions has the advantage of a clear iteration path defined through increasing the number of functions used. 
Generally a guess will not satisfy the original equations. 
However, by picking the coefficients of the linear combination correctly, it is possible to pick an optimal approximation to the correct answer. 
The idea of Galerkin is to fix these coefficients by adjusting them so that the guess satisfies the differential equations as a weighted average over some region of source space \cite{fletcher}.
The optimal definition of average is part of the problem of defining a rapidly convergent approximation scheme. 
Generally, the functions used to weight the average are chosen from the same set of complete functions as were used to build the guess. 
Careful implementation of this scheme allows the reduction of the error between the approximate solution  and the exact answer as the number of iterations is increased.

A crude, but very illustrative, way to calculate a field theory is to expand ${\cal Z}$ in a Taylor series in the sources and then truncate this series  at some maximal power \cite{sg1,john1,john2}. 
Such truncated series are definitely not solutions to non-trivial field theories. 
Nevertheless, it is reasonable to hope, for small sources, that the contributions from higher terms in the expansion rapidly decrease. 
We have checked and confirmed this on many problems where we know the answer or can check our results using Monte Carlo. 
The weight terms for the averaging procedure were usually chosen from the derivatives of the polynomial set used to approximate ${\cal Z}$. 
The accuracy that is achieved with a few terms is amazing and holds over a wide range of parameters.

Galerkin approximation is a weighted residual procedure. 
Such methods guarantee convergence in a mean. 
Our procedure is constructed so that the integral of the square of the local error vanishes in an appropriate limit.
The more familiar variational Hamiltonian approximation techniques are in the same general class of approximations. 
However, Source Galerkin methods do not have a direct correspondence to such methods as they have been used in high energy physics.

\section{Example application of Galerkin Method} 

Before we examine a field theory, we outline the application of Source Galerkin to a generic linear differential equation
\begin{equation}
\hat{L}f = 0.
\end{equation}
Here $\hat{L}$ is a differential operator defined in some domain $D$ with boundary conditions defined on a surface $\partial D$.  
We begin by guessing a solution
\begin{equation}
f^* = \varphi_0 \left( x_1,\ldots,x_N \right) + \sum_{j=1}^M a_j \varphi_j \left( x_1,\ldots, x_N \right).
\end{equation}
We assume the $\varphi_j$ are linearly independent and that they satisfy the appropriate boundary conditions but are otherwise unrestricted.

Now make the definition:
\begin{equation}
R \equiv \hat{L}f^* = \hat{L}\varphi_0 + \sum_{j=1}^M a_j \, \hat{L}\varphi_j.
\end{equation}
Next, pick $M$ linearly independent test functions $\rho_i \left( x_1,\ldots,x_N \right)$.
We do not confine our considerations to the usual Galerkin choice of picking $\rho_i = \varphi_i$. 
Fix the coefficients $a_1, \ldots, a_M$ by requiring
\begin{equation}
\langle R , \rho_i \rangle = 0, \quad i = 1, 2, \cdots, M.
\end{equation}
 We define this generalized dot product defined as follows:
\begin{equation}
\langle R, \rho_i \rangle \equiv \int \, dx_1\dots dx_N R( x_1,\ldots,x_N) \rho_i( x_1,\ldots,x_N) d( x_1,\ldots,x_N).
\end{equation}
To make sure that this is sensible,  $d$ is picked so that it dampens integrals.  
Options that we have used in field theory problems include restricting
 $d$ to be composed of  $\delta$ functions or choosing it to be of the form $\exp\left[-\sum_{i,j} x_{i} A_{ij} x_{j}\right]$.
After choosing $d$ and performing the above integrations, we obtain $M$ linear equations in $M$ unknowns.
If these are independent, we can solve for all of the unknowns. 
If not, we must change our choice of test functions or increase their number.
Under very general circumstances these approximations converge (weakly) to the exact answer as $M\to\infty$. 

\section{Fermions}

An exciting feature of our application of Galerkin methods to field theory is that fermions can be included in a natural manner.
Fermions present difficulty in Monte Carlo approaches for two reasons.
The first is because of the ``fermion determinant''. 
The problem occurs because after an approximate bosonic configuration is generated, the fermionic information contained in a determinant involving every point
of the lattice system must be evaluated. 
This evaluation requires a large amount of computer time and it must be done for many bosonic configurations. 
There are, of course, many schemes to decrease the amount of computation this requires. 
However, the difficulty is intrinsic to how path integration forces us to deal with fermions. 
Source Galerkin avoids this difficulty  because it allows making approximations for fermions without first solving a bosonic problem.
I will outline how this is done in the following.
The second problem is the well known ``fermion doubling problem''. 
This difficulty has nothing to do with numerical methods, but is an artifact
of lattice formulation. 
As mentioned, it is possible that we can avoid this as well. 
We have worked several fermionic problems in low dimensions on the continuum. 
This has been straightforward because the divergences we dealt with were trivial. 
We are currently working on a ``proof'' that Galerkin has enough flexibility to regularize all calculations without being forced to resort to a lattice.

The key difficulty in developing a Source Galerkin procedure for fermions is
that the na\"\i ve inner product definition is not appropriate because of the Grassmann nature of the sources. 
Direct integration of Grassmann polynomials is defined through the usual integration rules leads to a decimation of fermionic polynomials.
This makes it difficult to generate a formula to analyze
fermionic problems in an easy symmetric manner.
To avoid this, we define a similar inner product to that developed for bosons.
The definition we use is
\begin{equation}
\langle A [\bar{\eta} , \eta ] , B[ \bar{\eta} , \eta ] \rangle \equiv  \int [ d\bar{\eta} ] [d\eta ] \exp [ - \bar{\eta}\eta ] A [ \bar{\eta} , \eta ] B [
\bar{\eta} , \eta ].
\end{equation}

\section{Numerical Solutions} 

To test the source Galerkin ideas as applied to fermionic systems, we have studied simple quartic interactions on a one dimensional lattice.
These are described by the action:
\begin{multline}
{\cal S} = \sum_{i}^{N} \frac{1}{2} \bar{\psi}(i)  \left[  \psi(i + 1) - \psi(i - 1)\right]
 +  M \bar{\psi}(i) \psi(i)    \\
 +  \frac{G}{2} \bar{\psi}(i) \psi(i) \left[ \bar{\psi}(i+1) \psi(i+1)                                     + \bar{\psi}(i-1) \psi(i-1) \right] \\
 +  \left[ \bar{\eta}(i) \psi(i) + \bar{\psi}(i) \eta(i) \right].
\end{multline}

This action leads to the source functional differential equation:
\begin{multline}
\frac{1}{2} \left[ \frac{\partial{\cal Z}}{\partial {\bar{\eta}}(i+1)} - \frac{\partial{\cal Z}}{\partial {\bar{\eta}}(i-1)} \right]+ M \frac{\partial{\cal Z}}{\partial {\bar{\eta}}(1)} - {\eta}(1) {\cal Z} \\
- \frac{G}{2} \frac{\delta^3 {\cal Z}}{\delta {\bar{\eta}}(i)
                                   \delta {\eta}(i+1) \delta {\bar{\eta}}(i+1)}
 - \frac{G}{2}\frac{\delta^3 {\cal Z}}{\delta {\bar{\eta}}(i)
                                   \delta {\eta}(i-1) \delta {\bar{\eta}}(i-1)} = 0.
\end{multline}

If we restrict our study to 4 sites, this action can be solved exactly because of natural truncation of the power series expansion in the external sources at eighth order. 
Because we know the solutions, this model provides a non-trivial test case to examine the validity of our Galerkin methods.
We perform this test by assuming approximate solutions truncated at fourth and then sixth order in the fermionic sources. 
The weights of the polynomials in the sources are set by the Galerkin procedures outline above. 
To see exactly how these answers converge, we set $M=1$ and substitute a range
of values for $G$ and examine the behavior of the coefficients.
We find that the convergence is rather good. This is shown in Tables 1 thru 3
where we display the values for several couplings for the three independent values of the two-point Green's function.

\begin{table}
\begin{center}
\begin{tabular}{|llll|} \hline
\hfil g& \hfil 4th Order&\hfil 6th Order&\hfil Exact\\ \hline
\   \phantom{1}0     &\  0.75       &\  0.75       &\  0.75     \\ \hline
\   \phantom{1}0.1   &\  0.7003     &\  0.7089     &\  0.7095   \\ \hline
\   \phantom{1}0.5   &\  0.5553     &\  0.5840     &\  0.5926   \\ \hline
\   \phantom{1}1.0   &\  0.4428     &\  0.4803     &\  0.5      \\ \hline
\   10     &\  0.0996     &\  0.1160     &\  0.1494   \\ \hline
\end{tabular}
\end{center}

\caption{Convergence of fermionic Galerkin method
for $\langle \bar{\psi}(1) \psi(1) \rangle$}
\end{table}

\begin{table}
\begin{center}
\begin{tabular}{|llll|} \hline
\hfil   g     &\hfil 4th Order  &\hfil 6th Order  &\hfil  Exact  \\ \hline

\  \phantom{1}0     &\ 0.25       &\ 0.25       &\ 0.25    \\ \hline
\  \phantom{1}0.1   &\ 0.2413     &\ 0.2334    &\ 0.2328   \\ \hline
\  \phantom{1}0.5   &\ 0.2130     &\ 0.1928    &\ 0.1852   \\ \hline
\  \phantom{1}1.0   &\ 0.1871     &\ 0.1677    &\ 0.15     \\ \hline
\  10     &\ 0.0626     &\ 0.1141    &\ 0.0390   \\ \hline
\end{tabular}
\end{center}

\caption{Convergence of fermionic Galerkin method
for $\langle \bar{\psi}(2) \psi(1) \rangle$}
\end{table}

\begin{table}
\begin{center}
\begin{tabular}{|llll|} \hline
\hfil   g     &\hfil 4th Order  &\hfil 6th Order  &\hfil  Exact  \\ \hline

\  \phantom{1}0     &\ 0.25       &\ 0.25       &\ 0.25     \\ \hline
\  \phantom{1}0.1   &\ 0.2299     &\ 0.2223     &\ 0.2217   \\ \hline
\  \phantom{1}0.5   &\ 0.1726     &\ 0.1543     &\ 0.1481   \\ \hline
\  \phantom{1}1.0   &\ 0.1299     &\ 0.1118     &\ 0.1      \\ \hline
\  10     &\ 0.0201     &\ 0.0190     &\ 0.0065   \\ \hline
\end{tabular}
\end{center}

\caption{Convergence of fermionic Galerkin method
for $\langle \bar{\psi}(3) \psi(1) \rangle$}
\end{table}

While the 4 site problem is relatively straightforward, even a five site lattice is beyond simple direct solution.
However, the Galerkin results can be extended to a lattice of any size as long as the approximating polynomial is kept to lengths that are possible to handle in a computer. 
This is still a considerable restriction since the number of source polynomials eventually grows exponentially with the number of sites on the lattice. 
Nevertheless, if the higher order Green's functions of the theory are evenly moderately well behaved, we would expect an even limited expansion combined with Galerkin averaging to give a good answer for the lower order Green's functions. 
We have calculated the two point propagator for a eight site lattice using a Grassmann polynomial truncated first after fourth and then after sixth order. This is a non-trivial calculation since the fourth order polynomial by itself has has 124 coefficien
ts.
All coefficients were set by the Galerkin method.
These results are displayed in Table 4.

\begin{table}
\begin{center}
\begin{tabular}{|lll|}\hline
\hfil 4th Order &\hfil 6th Order  &\hfil Mean Field  \\ \hline
 0.5458    & 0.5831     & 0.6287       \\ \hline
 0.2288    & 0.2104     & 0.2218       \\ \hline
 0.0883    & 0.0753     & 0.0797       \\ \hline
 0.0177    & 0.0329     & 0.0243       \\ \hline
 0.0166    & 0.0267     & 0.0196       \\ \hline
\end{tabular}
\end{center}
\caption{Interacting Propagator on 8 Sites with $M=1.0$, $g=0.5$.}
\end{table}

In Table 4, we compare our results with fourth order and sixth order Galerkin on an eight site lattice to those obtained by a mean field calculation.
Mean field theory is a single pole approximation to the propagator.
The Source Galerkin calculated propagator, on the other hand, should include contributions from all poles. 
Nevertheless, for this problem the answers are very similar.
Calculations with the Source Galerkin method are very efficient, especially when compared to Monte Carlo.
Usually, for lattice gauge theory, the fermions must be quenched to make calculations tractable.
Here, determination of the interacting propagator for a system of dynamical fermions presents no special difficulty even in a pure fermionic theory.
The bulk of a calculation involves a single matrix inversion for a given set of parameters.
This is in contrast to Monte Carlo where many sweeps through the lattice are necessary to reduce statistical error.  
As our tables illustrate, the Source Galerkin calculations are very clean and show rapid convergence, when compared to the mean field results even at intermediate couplings and using only low order polynomials.
\bigskip
\section{Non-Linear Methods}
\medskip
Our initial examinations of the Galerkin method concentrated on
expansions that were simple for ${\cal Z}$. 
However, because of the disconnected Green's functions content of ${\cal Z}$, it is often much better to deal with $\ln{\cal Z}$. 
The difficulty with this is that the resulting equations are non-linear in the Galerkin parameters. 
For example, from the previous equations, it is straightforward to show that the approximation for a solution to quartic scalar field theory of the form
\begin{equation}
{\cal Z}^* = e^{J_{i} A_{ij} J_{j} + J_{l} J_{m} A_{lmpq} J_{p} J_{q} + \cdots}
\end{equation}
yields non-linear equations after being acted on by the source differential equations and then Galerkin averaged.  
These can be handled numerically. 
It is possible to develop iteration schemes based on the above guess which keep the problems associated with non-linearity under control.

An effective way to construct leading approximations is
to build ${\cal Z}$ in terms of spectral functions (propagators). 
A good combination can be guessed by writing down the perturbation theory or an effective theory for the action being studied. 
However, the only information we use from these structures is the general spectral forms. 
The masses and couplings are initially arbitrary and are set by Galerkin. 
Another effective and rapidly convergent guess can be made using polynomials weighted with propagators. 
There are a huge number of reasonable choices. 
In practice, we have found remarkable stability (per computational time) between very different types of iterations schemes when used on simple problems.

As a special case of the above non-linear form, we have been examining a spectral iteration scheme that we would think has much of the structure of an actual solution. 
The initial ans\"atz is taken to be roughly of the form
\begin{multline}
\ln {\cal Z} = \int \, dw \, dx {1\over 2} \, J (w) G_2^* (w,x) J
(x) \\ + 
\int \, dw\, dx\, dy\,dz\, {1\over 4} J (w) J(x) G_4^* (w,x,y,z)J(y) J(z) 
\end{multline}
with
\begin{equation}
\tilde{G}_2^* (p) = {1\over p^2+ \mu^2}
\end{equation}
and
\begin{equation}
 \tilde{G}_4^* ( p,q,r)  = \tilde{G}_2^* (p) \tilde{G}_2^* (q)
{B\over (p+q)^2 + \mu^{\prime 2} } 
 \tilde{G}_2^* (r) \tilde{G}_2^* (p +q - r ).
\end{equation}

We are generating results based on this starting form in 4 dimensions on fairly big systems. We are finding that this method consumes small amounts of CPU (relatively) but at the cost of large amounts of memory.  

\section{Boundary Conditions}

The discussion up to this point has avoided a fundamental issue. 
One of the most complicated and interesting problems associated with this type of numerical solution lies in the analysis of the appropriate boundary conditions imposed on the source differential equations. 
If we look at the special case of zero space time dimensions (ultralocal model) there is only one differential equation. 
It is a third order differential equation so its solution involves the specification of three {\it a priori} arbitrary constants \cite{cooper}.
In particular, the normalization of ${\cal Z}$, and the one and two field Green's functions are unspecified. 
The situation for arbitrary space-time dimension is more complex with a correspondingly greater number of unspecified constants.
This situation is not entirely unfamiliar. 
In this discussion we have excluded terms involving odd powers of the sources.
This constitutes a boundary condition on the first derivative of ${\cal Z}$. 
We know that, if we do not exclude odd terms, we can under the correct conditions write solutions that have spontaneous symmetry breaking. 
Dealing with the boundary condition on the second derivative terms is a much more complicated and a very interesting problem that leads to the enhanced understanding of phase structure mentioned earlier. 
In particular we find the possibility of solutions to quantum field theory which are extremely singular for small bare couplings.
We analyze this elsewhere \cite{zach1,zach2,gg1}. 
I have avoided direct contact with this problem in the example detailed here by truncating the Taylor series solution in ${\cal Z}$. 
It is easy to see that truncation forces the approximation to agree with the smooth $g=0$ limit of the theory that corresponds to perturbation theory. 
Therefore, truncation imposes an implicit boundary condition on the second derivative terms.

\section{Conclusions}

Our previous published work in developing the Source Galerkin method has been focussed
primarily on scalar $\phi^4$ in various spacetime dimensions 
and lower dimensional lattice four Fermi
theories \cite{sg1,john1,john2}. 
Now that we have an understanding of these systems we have begun to analyze gauge and fermionic theories in four dimensions.
We have developed more sophisticated expansion that the direct series methods and have a good understanding of how to use Galerkin methods to set parameters in effective field theories. 
We believe that Galerkin techniques can be effectively used on the continuum and have verified this in two dimensional models. 
We have developed what appears to be a correct working procedure to deal with continuum four dimensional problems.

As we develop our techniques we anticipate to be able to answer questions about more realistic theories. 
At this stage, I am very optimistic that the Source Galerkin model will be important in making a contribution to our understanding of the solutions of quantum field theory.

\section*{Acknowledgment}

Results presented here were obtained in collaborations with S. Garc\'ia (IBM), Z. Guralnik (Princeton), S. Hahn (Brown) and J. Lawson (ICTP Trieste).

\end{document}